\def\beq{ \begin{equation}}
\def\eeq{ \end{equation}}
\def\ba{ \begin{eqnarray}}
\def\ea{ \end{eqnarray}}

\def\half {{1\over 2}}
\def\omicron {{\rm o}}
\documentstyle[aps,preprint,12pt]{revtex}
\tightenlines
\title{ Radiation Reaction fields for an accelerated dipole
 for scalar and electromagnetic radiation}

\author{ W. Unruh}
\address{ Program in Cosmology and Gravity of CIAR\\Dept 
Physics and Astronomy\\University 
of B.C.\\ Vancouver, Canada V6T 1Z1}
\begin{document}
\maketitle

\begin{abstract}
The radiation reaction fields are calculated for an accelerated changing
dipole in scalar and electromagnetic radiation fields. The acceleration
reaction is shown to alter the damping of a time varying dipole in the
EM case, but not the scalar case. In  the EM case, the dipole radiation
reactionfield can exert a force on an accelerated monopole charge
associated
with the accelerated dipole. The radiation reaction of an accelerated
charge
does not exert a torque on an accelerated magnetic dipole, but an
accelerated dipole does exert
 a force on the charge.  The technique used is that originally
developed by Penrose for non-singular fields and extended
 by the author for an accelerated monopole charge.
\end{abstract}

It is well known that an accelerated charge radiates and that the
 emitted field from the accelerated charge exerts a force ( the
 radiation reaction force) back on the charge itself. This radiation
reaction force is
 usually derived either by an appeal to a balance of the energy and
 momentum emitted by the charge or by a detailed examination of the
energy-momentum
 tensor just near the charge. These techniques invariably have
 difficulties with the fact that the fields from a point charge
 diverge, and necessitate re-normalisation of various quantities (eg
 the mass) in order to extract reasonable results. However, it was
discovered by Penrose and Unruh that the
 radiation reaction field for a charge could be extracted from the
 radiation field by means of an integral over the future  null cone of
the the particle.
 This integral, which in the absence of any sources, exactly gives the
 field strength at a point, also gives a finite result whan applied to
the field emanating from a point charge, and that finite result is
exactly the radiation reaction fields at the particle.  In this paper I
will show that this approach
 can be generalised to the case of an accelerating dipole and leads to
 finite radiation reaction fields at the location of the point dipole.
 These fields lead to the damping forces on that accelerating dipole.
These results will ultimately be applied in another paper on the
question of the equilibrium polarisation of an accelerating particle
with spin, but seem to be of
 sufficient general interest that they are here separated out for
 detailed study.

In this paper, I will derive the radiation reaction fields at the
location of point source for a massless scalar field, and the reaction
fields at the location of
 a point dipole source for the electromagnetic field, using the Penrose
 integral to do so. Thus I will begin by giving a very brief review of
the Newmann-Penrose (NP)
 spinor formalism, and the natural null tetrad metric for an
 accelerated path in Minkowski spacetime. That accelerated path is
assumed for the purposes of this paper to consist of an acceleration
restricted to a plane (linear or circular acceleration,
 but with non-constant acceleration).

\section{Null Metric, tetrad and NP spin formalism}

This section will be a very quick review of the NP spinor formalism,
presented as much to specify the notation I will use as for any other
purpose. For a more complete introduction to spinors, see Penrose and
Rindler \cite{PR}

Given a path in spacetime, define a $ur\theta\phi$ coordinate system as
follows:

Parameterize the curve by its length parameter $u$.  (ie,
$\eta_{\mu\nu}{dx^\mu\over du}{dX^{\nu}\over du} =1)$. At each point
along the curve,
 define the future directed null cone centered at that point along the
 curve, and label that null cone by the parameter $u$.

On each of the null cones, choose the direction of the acceleration
vector, and call it the $\theta=0$ direction. Along each of the null
generators of the cone (null lines which originate
 at the point along the curve at the vertex of the cone) define a
 radial affine parameter $r$, such that the tangent vector $n^\mu$ to
 the null curve parameterized by r (which is a null vector) has a dot
product with the tangent vector to the original curve of unity--
Ie, $n^\mu {dx_\mu\over du} = 1$. Now, the two spheres defined by $u $
and $r$ constant are
 metric two spheres on which we will define angular coordinates $\theta
 $ and $\phi$. They are chosen so that the metric on these two spheres
is the usual two sphere metrics.. Because of the way that r has been
defined, the circumference of these two spheres will be $r$ so the
metric on these two spheres will be $r^2(d\theta^2+d\phi^2)$. The
direction $\theta=0$ is at the null generator pointing in the
 direction of the acceleration vector, and $\theta$ and $\phi$ will
 label the null generators. 
This procedure will define a new coordinate system centered on the path
of the particle. The metric of flat spacetime in these coordinates  is
\ba
 ds^2 &= &\left(1+2 f(u) r \cos(\theta)-r^2(g(u) \cos(\phi)   
+ f(u) \sin(\theta))^2  -r^2 g(u)^2 \sin(\phi)^2\cos(\theta)^2 \right)
d  u^2 
\nonumber 
\\
&&+2   d  u  d   r   +2 (f(u) \sin(\theta)+g(u) \cos(\phi)) 
r^2   d  u  d \theta-2 g(u) \sin(\theta) \cos(\theta) \sin(\phi) 
r^2   d  u  d   \phi 
\nonumber \\
&&-r^2   d  \theta^2  -r^2 \sin(\theta)^2   d  \phi^2  
\ea
where $f(u)$ and $g(u) $ will be related to the acceleration. $f(u)$ is
the acceleration, and $g(u)$ is the rate of change of the direction of
the acceleration. The acceleration is assumed
 to be confined to the place $\phi=0$ or  $ \pi$.

We now define a null complex tetrad for this metric. This is a set of
four vectors $l^\mu$,
 $n^\mu$, $m^\mu$ and $\bar m^\mu$ such that each of these vectors is
 null (has zero inner product with itself) and such that
\beq
l^\mu n_\mu=m^\mu\bar m _\mu=1
\eeq
and all other inner products are zero.  The vector $n^\mu$ has already
been defined as the vector tangent to the null generators of the null
cones. The vector $m^\mu$ is defined as a complex vector lying tangent
to the two surface of the $u$ and $r$ constant spheres. $\bar m^\mu$
is  the complex conjugate to
 $\mu$. Thus the null  vector $l^\mu$ will be orthogonal to the these
 two spheres (as is $n^\mu$).  $l^\mu$ is assumed to be such that
 $l^\mu {dX_\mu\over du}$ is greater than zero (ie, is a future
pointing null vector.) For definiteness, I will choose $m^\mu=
e_\theta^\mu +ie_\phi^\mu$ where $e_\theta$ and $e_\phi$ are the unit
vectors
 which lie along the $u,~r,~\phi$ constant line and $u,~r,~\theta$
 constant lines respectively.

In the above $ur\theta\phi$ coordinate system, these vectors therefor are
\ba
n^\mu&& = \left[0,1,0,0\right]\\
m^\mu&&= {1\over \sqrt{2}}\left[0,0,{1\over r},
 {i\over r \sin(\theta)}\right]\\
\bar m^\mu&&= {1\over \sqrt{2}} \left[0,0,{1\over r}, 
{-i\over r \sin(\theta)}\right]\\
l^\mu&&= [1, -{1\over 2}-f(u) r \cos(\theta) ,
{}~ f(u) \sin(\theta)+g(u) \cos(\phi),
{} ~-g(u) \cot(\theta) \sin(\phi)] 
\ea

In addition to the null vectors, the formalism defines a set of two
dimensional complex spinors. In particular they assumed that there are
two separate two dimensional spinor spaces which are anti-unitarily
related to each other (these are just the two unitarily inequivalent
spin 1/2 representations of the Lorentz group). Tensors over these two
spaces are designated
 by indexed symbols whose indices are capital Roman letters. Tensors
 over the one representation have plain indices and tensors over the
other are designated with primes on their indices. Because of their
anti-unitary relationship, there exists a mapping from one type of
tensor to the other, which we will denote by complex conjugation. Thus
$(w^A)^*= \bar w^{A'}$. This is defined so that the inner product of
complex conjugate vectors is just the ordinary complex conjugation.
$(w^A v_A)^*=\bar w^{A'} \bar v_{A'}$.
 A complete set of basis vectors on these spinor spaces are designated
 by $\iota^A,~\omicron^A$ in the one case, and $\iota^{A'},~
\omicron^{A'}$ in the other case.They are chosen so that
$(\iota^A)^*=\iota^{A'}$, etc.  These spinor spaces are
 related to the spacetime vectors by means of the spin matrices
 $\sigma^\mu_{AA'}$, matrices which for a given vector $v_\mu$
 represent a mapping from the one spinor space to the other via
$V_\mu\sigma^\mu_{AA'}$. If we choose the basis vectors appropriately
 in the spinor space, then these matrices are just the four Pauli spin
 matrices,${\bf 1},\sigma_x,\sigma_y,\sigma_z$. In particular I will
 assume that they are chosen so that
\ba
l_\mu\sigma^\mu= ({\bf 1} + \sigma_z)/2\\
n_\mu\sigma^\mu=({\bf 1} - \sigma_z)/2\\
m_\mu\sigma^\mu=(\sigma_x+i*\sigma_y)/2\\
\bar m_\mu\sigma^\mu=\sigma_x-i*\sigma_y)/2
\ea
with $\iota^A$ and $\iota_{A'}$ both represented by $(0,1)$ and
$\omicron_A$ and $\omicron _{A'}$
 both represented by $(1,0)$. Thus, we have
\ba
l_\mu\sigma^\mu_{A A'} &&= \omicron_A\omicron_{A'}\\
n_\mu\sigma_{AA'}&&=\iota_A\iota_{A'}\\
m_\mu\sigma^\mu_{A A'}&&=\omicron_A\iota_{A'}\\
\bar m_\mu\sigma^\mu_{A A'}&&=\iota_A\omicron_{A'}
\ea

From now on I will freely alternate between the spinor representation
and the vector
 representation. Thus a tensor $S_{\mu\nu}$ can also be written as
 $S_{AA'BB'}=  S_{\mu\nu}\sigma^\mu_{A A'}\sigma^\nu_{BB'}$, etc.

These spinor spaces also have a metric defined on them, a metric which
must be compatible
 with the Lorentzian metric. The metric, designated by either
 $\epsilon^{AB}$ for the spinor, or $\epsilon^{A'B'}$ for the
anti-spinor space is an antisymmetric metric, such that
$\epsilon^{AB}=-\epsilon^{BA}$. The inverse metric is $\epsilon_{AB}$
which is also antisymmetric and is chosen so that
$\epsilon_{AB}\epsilon^{CB}=\delta_{A}^{C}$.
 Indices are raised and lowered by means of the metric, but because of
 the antisymmetry of the metric, the order of the indices is of crucial
 importance. My convention, following Penrose and Rindler\cite{PR}, is
\ba
\omega^A = \epsilon^{AB}\omega_{B}=-\omega_B \epsilon^{BA}\\
\omega_A= \omega^B \epsilon_{BA}.
\ea
and similarly for the prime space.  The basis vectors $\iota^A$ and
$\omicron^A$ are chosen to obey $\omicron_A\iota^A=1$, which is
compatible with the  relation of these spinor bases to the spacetime
null tetrad, and the normalisation of those tetrad vectors.

 This antisymmetry of the metric and the attendant risk of confusion in
 index manipulations (together with the presence of the two separate
 types of spinors) is probably the greatest impediment to the adoption
of spinors as a standard approach in special and general relativity.
Given these disadvantages, there must be some advantages to the spinor
formalism which would persuade any but masochists to adopt the
formalism. The key advantages are firstly that spinors allows a
unified  treatment of the various fields
 of whatever spin type  in one simple notational system. Ie, scalar,
 spin 1/2, vector, spin 3/2, .... fields can all be treated very
 similarly,  in an extremely compact and transparent way. The second
advantage is that spinor space is a two dimensional space. This means
that there is only antisymmetric tensor of rank two, and all other
antisymmetric tensors of rank two must be proportional to this tensor.
Furthermore, since the metric is antisymmetric, it can be chosen as the
fiducial antisymmetric tensor.  Any two indices (of the same type) of a
spinor tensor can always be written as a combination of symmetric and
antisymmetric pairs.  Thus, we can write any tensor
\beq 
S^{...A..B..}_{...} = S^{...(A|..|B)..}_{...}+ \half\epsilon_{CD}
 S^{....C..D..}_{...} \epsilon^{AB}
\eeq
Any tensor can thus be written as the sum of tensors which are totally
symmetric on their indices times products of the metric tensors.  This
ability to represent any tensor as either entirely symmetric tensors or
multiples of
 the metric is the key power of the spinor notation, and achieves its
 greatest power in the representation of massless fields. A massless
field of spin $s$ is represented by a tensor with $s$ indices, all of
which are completely symmetric, say $\Psi_{AB..S}$ where
 the tensor is symmetric under interchange of any two indices.
 Furthermore, the equations of motion of a spin $s$ massless field are
simply written as $\nabla^{AA'}\Psi_{AB...S}=0$, where $\nabla_{AA'} $
is the covariant derivative, defined on spinors such that
$\sigma^\mu_{AA'}$ and the metric $\epsilon_{AB}$ are covariantly
constant. It is the compactness of the spinor notation, and the
transparency of the symmetries of the fundamental tensors which give
the spinor notation its power.

In the following I will be primarily interested in the electromagnetic
field, $F^{\mu\nu}$.  Writing this in spinor form, we have $F_{AA'BB'}$
with the antisymmetry ensuring that $F_{AA'BB'}=-F_{BB'AA'}$. But using
the above reduction, we note that this can be written as
\beq
F_{AB}=\half(\epsilon^{C'D'}F_{A C'BD'}\epsilon_{A'B'}
+\epsilon^{CD}F_{CA'DB'}\epsilon_{AB})
\eeq
I will use the notation that 
 \beq
F_{AB}=\half \epsilon^{C'D'}F_{A C'BD'} 
\eeq
so that
\beq
F_{A A' B B'}=F_{AB}\epsilon_{A'B'}+F_{A'B'}\epsilon_{AB}
\eeq
   Maxwell's equations become   
\ba
\nabla^{AA'}F_{AB}=4\pi J_{B}~^{A'}\\
\nabla^{AA'}F_{A'B'}=4\pi J^{A}~_{B'}
\ea
where $J^\mu$ is the current source for the Maxwell field. 
The reality of $F_{\mu\nu}$ and $J_\mu$ ensure that
 $(F_{AB})^*=F_{A'B'}$ and $(J_{AB'})^*=J_{BA'}$.

Finally, if $G_{\mu\nu}$ is another antisymmetric tensor, then
\beq
F^{\mu\nu} G_{\mu\nu}=2(F^{AB} G_{AB} + F^{A'B'} G_{A'B'})
\eeq
and if $F$ and $G$ are both real tensors, then 
\beq
F^{AB} G_{AB} = ( F^{A'B'} G_{A'B'})^*
\eeq

Having established the notation, I will now state the theorem without
proof\cite{Unruh,PR}. Given a spin $s$ massless field without source,
then the following integral gives the value of the field
$\Psi(0)_{AB...S}$ at the apex (r=0) of the null cone emanating from a
point in spacetime.

\ba
T(0)^{AB..S}\Psi(0)_{AB..S} =
 {(-1)^{2s+1}\over 2\pi} \int_{r,u{\rm ~const}} r \sin(\theta) 
&&T_{AB..S}\iota^A\iota^B...\iota^S  \omicron^{D}\omicron^E...\omicron^T
\left(\omicron^X\omicron^{Y'} \nabla_{XY'}\Psi_{DE..T}\right.\\
&&\left. -(2s+1)\Psi_{DE..T}\omicron^W \iota^X\omicron^{Y'}
\nabla_{XY'}\omicron_W\right)d\theta d\phi
\nonumber 
\ea
Here $T_{AB..S}$ is any covariantly constant spinor field,
$\nabla_{XX'}T_{AB..S}=0$. (In
 fact it need only be covariantly constant along the null cone of
 interest).   This theorem states that the value of any massless field
 in flat spacetime can be determined by the integral over a sphere on
the null cone emanating from that point. This expression is a
generalisation to a dynamic massless field of arbitrary spin  of the
Kirkoff type integrals for static fields in terms of the integral over
some surface surrounding the point in question of the normal
derivatives of the field and the Green's function for that field.

Although I have stated the theorem in terms of integrals over metric
spheres on the null cone ($r=$const.), it can also be generalised to
the integral over arbitrary two surfaces on the null cone. However, I
will not use that generalisation here.

While the use of this integral for the value of the field at points in
the spacetime where the field is regular is interesting but
unexceptional, a surprising result \cite{Unruh} is that this integral
also gives finite values if the point of interest is the location of a
point charge (with
 its divergent Coulomb field). In fact, this integral (or rather the
 average of this integral over the future null cone emanating from the
location of that charge at some time, and the past directed null cone
emanating from that same point) gives exactly the radiation reaction
field for an accelerating charge.
\beq
F_{RR}^{\mu\nu}= -{4\over 3}{dx(u)^{[\mu}\over du}
{D^3   x(u)^{\nu]}\over Du^3}
\eeq
where the square brackets around the indices indicates
anti-symmetrization, ($S^{[\mu\nu]}=\half(S^{\mu\nu}-S^{\nu\mu})$.
Ie, this Penrose integral automatically averages out the divergent
field of the point particle to give just the finite radiation reaction
contribution. Thus this Penrose integral approach differs substantially
from techniques like the Dirac\cite{Dirac} or
Abraham--Lorentz\cite{Jackson}   which   suffer from divergences and
the necessity for re-normalisations.

The purpose of this paper will be to apply the above formula to the
calculation of the radiation reaction field of an accelerating and time
varying point dipole source for the
 electromagnetic field. Surprisingly, considering the fact the
 "Coulomb" portion of the field now diverges as $1/r^3$ rather than as
$1/r^2$ for the case of a point charge, the Penrose integral is still
finite, and gives a field which agrees with the radiation reaction
field for a time varying unaccelerated dipole calculated by other
methods\cite{dipole}. I will therefor assume that this finite field is
also   the correct radiation reaction field for an accelerated dipole.
I have so far been unable to prove that this is consistent with the
radiation damping one would calculate by more traditional techniques
(eg, from the energy momentum tensor), but I can see no reason why it
would not.

\section{Scalar Field}

In order to gain practice, let me first calculate the radiation
reaction field produced by
 a source which radiates scalar radiation. Consider a scalar field
 coupled to a point source travelling along the line $x^{\mu}(u)$ in
flat spacetime, with a time varying source for the scalar field of
intensity $m(u)$, The Lagrangian for the massless scalar field is
assumed to be
\beq
S= \int \half \sqrt{-g} \phi_{,\mu}\phi_{\nu}d^3 x 
+ 4\pi \int m(u)\phi({\bf x}(u))du
\eeq
which gives the equation for $\phi$ of 
\beq
\nabla^\mu\nabla_\mu \phi = 4\pi \int m(u)\delta^4(x^\mu-x(u)^\mu(u)) du
\eeq
In the $ur\theta\phi$ coordinates, the retarded Green's function for
the scalar field in $ur\theta\phi$ coordinates is particularly simple,
it is just $1/r$. Thus the retarded solution for the scalar field is
just
\beq
\phi(u,r,\theta\phi)={m(u)\over  r}
\eeq
We can now substitute this expression into the equation for the
radiation reaction field.
 In order to write this in a slightly more transparent form, recall
 that $l^{A A'}=\omicron^A\omicron^{A'}$. Thus
\ba
\omicron^X\omicron^{Y'}\nabla \phi &&= l^\mu\phi_{,\mu}\\
\omicron _A \omicron^X\omicron^{Y'}\nabla_{XY'}\omicron^A&&=
 \omicron _A \omicron_{A'} \omicron^X\omicron^{Y'}
\nabla_{XY'}(\omicron^A\iota^{A'})
\nonumber \\
&&=l_\mu m^\alpha\nabla_\alpha m^\mu
\ea
since $\omicron_A\omicron^A=0$. 
This gives us, 
\ba
\phi(0)&&=-{1\over 2\pi} \int r \sin(\theta)\left( {\dot m(u)\over r} 
-(\half+r f(u)\cos(\theta))(-{m(u)\over r^2}) 
-\half {m(u)\over r^2}\right) d\theta d\phi
\nonumber \\
&&= -2 \dot m(u)
\ea

An interesting application is where the source for the scalar field is
an internal oscillator with configuration variable $q$, such that
$m=\epsilon q$. Ie, the Lagrangian is
\beq
\int \sqrt(-g)\phi_{,\mu}\phi_{,\nu}g^{\mu\nu} d^4x 
+ \int\left( \half (q_{,u}^2 -\Omega^2 q^2) +\epsilon q \phi(x(u))\right) du
\eeq
The equation of motion for the oscillator, including the effect of the
radiation reaction field, is now
\beq
-q_{,u,u}-\Omega^2 q +\epsilon( \phi_0(x(u)) -2\epsilon q_{,u})=0
\eeq
where $\phi_0$ is the value of the background field at the location of
the particle.  Ie, the radiation reaction field acts as a simple
damping term to the internal harmonic oscillator, with damping
coefficient $2\epsilon^2$. In another paper I will use this to
investigate the emission of radiation from an accelerating detector in
interaction with the scalar field.

\section{Electromagnetic radiation reaction fields}

To find the radiation reaction fields for the electromagnetic field, we
must first solve the equations for the electromagnetic fields from a
point dipole. First define the vector tangent to the path of the
particle parallel transported over the null cone.
\beq
T^\mu= l^\mu+\half n^\mu
\eeq
In addition, define the vectors 
\ba
Z^\mu&&= (-l^\mu +\half n^\mu)\cos(\theta) 
- {1\over \sqrt{2}}\sin(\theta) (m^\mu+\bar m^\mu)\\
X^\mu&&= \left((-l^\mu +\half n^\mu)sin(\theta) + 
{1\over \sqrt{2}}\cos(\theta) (m^\mu+\bar m^\mu)\right)\cos(\phi) + 
{i\over \sqrt{2}} (m^\mu-\bar m^\mu) \sin(\phi)\\
 Y^\mu&&= \left((-l^\mu +\half n^\mu)\sin(\theta) +
 {1\over \sqrt{2}}\cos(\theta) (m^\mu+\bar m^\mu)\right)\sin(\phi) -
 {i\over \sqrt{2}} (m^\mu-\bar m^\mu) \cos(\phi)
\ea
which are all vectors which are parallel over the whole of the surface
of the cone $u$ constant (but are not parallel off that cone.) For
future needs, let me define the basis vectors
$e^{(i)\mu}$ such that 
\ba
e^{(0) \mu}=T^\mu\\
e^{(1) \mu}=X^\mu\\ 
e^{(2) \mu}=Y^\mu\\
e^{(3) \mu}=Z^\mu
\ea
Define the dipole moment
 \beq
{\cal D}^\mu (u)= d_x(u)X^\mu +d_y(u) Y^\mu 
+d_z(u) Z^\mu=d_{(i)}e^{(i)\mu}
\eeq
as the dipole moment vector, with $d_{(0)}$   zero. Furthermore, define 
\beq
S^{\mu\nu}= {1\over r}(T^\mu {\cal D}^\nu-T^\nu {\cal  D}^\mu)
\eeq
Then the vector potential for the electric dipole moment ${\cal D}$ is
\beq
A^\mu(u,r,\theta,\phi)= \nabla_\nu S^{\mu\nu}
\eeq
with electromagnetic field
\beq
F_{\mu\nu}= \nabla_\nu A_\mu-\nabla_\mu A_\nu
\eeq
Also define the tensor
\beq
{\cal A}^{\mu\nu}=a_{(ij)}e^{(i)\mu}e^{(j)^\nu} 
\eeq
where $a_{(ij)}$ is antisymmetric in $ij$. 
Then the Penrose integral equation becomes
\ba
{\cal A}^{\mu\nu}F_{RR\mu\nu}&&=
-\half{1\over 2 \pi} \left[ \int r \sin(\theta) 
{\cal A}_{AB}\iota^A\iota^B \omicron_C\omicron_D
\left(\omicron^X\omicron^{Y'}\nabla_{XY'} F_{CD} \right.\right.
  \\
&&~~~~~~~~~~~~~~~~~~~~~  \left.\left. -
 3 F_{CD}\omicron^W \iota^X\omicron^{X'} \nabla_{XX'} 
\omicron_W \right)d\theta d\phi\right] 
+ ComplexConjugate
\nonumber \\
&&=-{1\over 4 \pi} \int r \sin(\theta){\cal A}_{\mu\nu}
\bar m^\mu n^\nu l^\rho m^\sigma( l^\tau\nabla_\tau F_{\rho\sigma}
- 3 F_{\rho\sigma} m^\alpha \bar m^\beta \nabla_\beta l_\alpha ) 
d\theta d\phi +CC
\nonumber 
\ea
The first factor of half arises from the averaging over the future and
past null cones (the contribution from the past null cone being zero).
 
This integral, though very messy, can be evaluated, and gives the
radiation reaction field. After extensive  calculation, aided in an
essential way with the GRTensorII computer algebra system\cite{grii},
the result for the $E$ and $B$ fields are
\ba
E^{(i)}_{RR}=-{2\over 3} {D^3_{FW} {\cal D}^{(i)} \over Du^3}
- {D_{FW}{\cal D}^{(i)}\over Du}f^2(u) + {\cal D}^{(i)} f(u)\dot f(u) \\
B^{(i)}_{RR}={1\over 3}\left( {D^2_{FW}a_{(j)}\over Du^2}{\cal D}_{(k)}
-2 {D_{FW}a_{(j)}\over Du}{D_{FW}{\cal D}_{(k)}\over Du} \right)
\epsilon^{(i)(j)(k)}
\ea
where ${D_{FW}\over Du}$ is the Fermi Walker derivative of the quantity
 along the path of the particle 
\beq
{D_{FW}S^\mu\over Du}= T^\nu\nabla_\nu S^\alpha(\delta^\mu_\alpha 
-T^\mu T_\alpha).
\eeq
and
\beq
a^{(i)}=e^{(i)}_\mu T^\nu \nabla_{\nu}T^{\mu}
\eeq
 are the components of the acceleration in the $XYZ$ frame.

We can find the radiation reaction fields for the case for a magnetic
dipole moment, ${\cal M}^\mu$ by simply taking the dual of all of the
above equations, for which we get
\ba
B^{(i)}_{RR}=-{2\over 3} {D^3_{FW} {\cal M}^{(i)} \over Du^3}
- {D_{FW}{\cal M}^{(i)}\over Du}f^2(u) 
+ {\cal M}^{(i)} f(u)\dot f(u)\\
E^{(i)}_{RR}=-{1\over 3}\left( {D^2_{FW}a_{(j)}\over Du^2}{\cal M}_{(k)}
-2 {D_{FW}a_{(j)}\over Du}{D_{FW}{\cal M}_{(k)}\over Du} \right)
\epsilon^{(i)(j)(k)}
\ea

We note that if the acceleration is not equal to zero, and the dipole
moment is periodic
 in time, the acceleration increases the radiation reaction field,
 since $D^3_{FW}{M^{(i)} \over Du^3}\approx -Omega^2 { D_{FW}M^{(i)}
\over Du}$, unlike for the scalar field where the radiation reaction
field was independent of the acceleration.

One of the most interesting features of the radiation reaction field is
that an
 accelerated magnetic dipole will have an electric component to its
 radiation reaction field. Thus, if the point dipole also has a
non-zero charge, then the accelerated dipole will exert a force on that
accelerated charge in addition to the normal radiation reaction force
from the charge itself. On the other hand, since the radiation reaction
field for an accelerated charge has no magnetic component in the rest
frame of that charge, the accelerated charge will not alter the motion
of the magnetic dipole moment.

If the magnetic dipole moment is proportional to the angular momentum,
via the magnetic moment, $\mu$ of the point charge distribution, then
the equation of motion for the dipole moment under constant magnitude
of acceleration is
\beq
{D_{FW}{\bf M}\over Du}= \mu{\bf M}\times{\bf B}
 = \mu{\bf M}\times (\bf B_0+B_{RR})
\eeq

 This solution is subject to runaway behaviour just as is the
 accelerated charge. In this case the runaway behaviour is in the
 direction in which the angular momentum points, the angular momentum
 itself of course being conserved by the above equation. Choosing
 coordinates $\theta\phi$ such that the ${B_0}$ points in the direction
 of $\theta=0$, and the magnetic moment vector points in the direction
 $\Theta(u),~\Phi(u)$, and choosing $g(u)=0,~F(u)$ constant,  the
 equation of motion for $\Theta,\Phi$ are
\ba
\Theta_{,u}&&= -{2\mu^2 \over 3} \left( \Phi_{,uuu} \sin(\Theta(u))
+3\left(\Phi_{,u} (\sin(\Theta(u)))_{,u}\right)_{,u}
-\sin(\Theta)(\Phi_{,u}^3 +f^2 \Phi_{,u})\right)\\
\Phi(u)_{,u}&&= -\mu B_0 -{2\mu^2\over 3\sin(\Theta(u))}
\left( \Theta_{,uuu} - 3\cos(\Theta(u)) (\sin(\Theta(u))
\Phi_{,u})_{,u}-(\Theta_{,u})^3 +f^2 \Theta_{,u})\right)
\ea
Using the lowest order solution for $\Phi(u)$, namely $\Phi_{,u}= -\mu
B_0$, and neglecting all but the lowest order terms in the equation for
$\Theta$, we get
\beq
\Theta_{,u}= \sin(\Theta){2\mu^2\over 3}
\left( (\mu B_0)^2 +f^2)\mu B_0\right)
\eeq
which has solution
\beq
\cos(\Theta)= -tanh \left({2\mu^2\over 3}( (\mu B_0)^2 
+f^2)\mu B_0(t-t_0)\right)
\eeq
 However, retaining the higher order terms  leads to  runaway
 solutions, where in particular $\Phi$ diverges exponentially. Ie, as
 usual, the electromagnetic radiation reaction is useful only in
 providing the lowest order corrections to the solution, and cannot be
 taken seriously as a complete solution.

\section*{Acknowledgements}
I would like to thank P. Chen for inviting me to the Quantum Aspects of
Beam Physics Workshop and J.D. Jackson for raising the issue of the
radiation for circular accelerating beams and thus reviving my interest
in this problem. I would also like to thank the Canadian Institute for
Advanced Research and NSERC for support during this research.

\end{document}